\documentstyle[aps,preprint,epsf]{revtex}
\begin{document}
\draft

\title{Quantum chaos in multicharged ions and statistical approach to
the calculation of electron-ion resonant radiative recombination}

\author{G. F. Gribakin, A. A. Gribakina, and V. V. Flambaum}

\address{School of Physics, The University of New South Wales,
Sydney 2052, Australia}

\date{\today}

\maketitle

\tightenlines

\begin{abstract}

We show that the spectrum and eigenstates of open-shell multicharged
atomic ions near the ionization threshold are chaotic, as a result of
extremely high level densities of multiply excited electron states
($10^3$ eV$^{-1}$ in Au$^{24+}$) and strong configuration mixing. This
complexity enables one to use statistical methods to analyse the system.
We examine the dependence of the orbital occupation numbers and
single-particle energies on the excitation energy of the system, and show
that the occupation numbers are described by the Fermi-Dirac distribution,
and temperature and chemical potential can be introduced. The Fermi-Dirac
temperature is close to the temperature defined through the canonical
distribution. Using a statistical approach we estimate the contribution
of multielectron resonant states to the radiative capture of low-energy
electrons by Au$^{25+}$ and demonstrate that this mechanism fully accounts
for the $10^2$ times enhancement of the recombination over the direct
radiative recombination, in agreement with recent experimental observations.

\end{abstract}

\vspace{18pt}

%****************************************************************************

\section{Introduction}

In this paper we investigate the spectrum and eigenstates of a multicharged
positive ion at energies close to its ionization threshold $I$. Using
Au$^{24+}$ ($I=750$ eV) as an example, we show that this spectrum is
dominated by multiple electron excitations into a few low-lying unoccupied
orbitals. As a result, it is extremely dense, with level spacings
$\sim 1$ meV between the states of a given total angular momentum and parity
$J^\pi $. The electron Coulomb interaction induces strong mixing of the
multiply-excited configurations, which leads to a statistical equilibrium
in the system. The latter is similar to a thermal equilibrium, and
variables such as temperature can be introduced to describe it. This enables
one to use a statistical approach in the situation where a full dynamical
quantum calculation is simply impossible because of the enormous size of the
Hilbert space ($\gtrsim 10^5$ for Au$^{24+}$).

We apply this approach to the problem of radiative capture of low-energy
electrons by multicharged positive ions, and show that in these systems the
contribution of resonant {\em multielectronic} recombination that proceeds
via electron capture into the {\em multiply-excited} compound states, is
responsible for high recombination rates, much greater than those expected
from the simple direct radiative recombination. Our calculation resolves
quantitatively the long-standing puzzle of huge enhancements of the
electron-ion recombination rates, and essentially removes the
``enormous discrepancies between theoretical and experimental rate
coefficients'' (Hoffknecht {\em et al.} 1998). The situation here turns out
to be similar to the radiative neutron capture by complex nuclei [$(n,\gamma )$
reaction] where the resonance mechanism involving the compound nucleus states
is also much stronger than the direct capture (Flambaum and Sushkov 1984,
1985).

So far the enhancement of the recombination rates at low electron energies
$\lesssim 1$ eV has been observed for a number of ions\footnote{Apart from
the enhancement at eV energies due to many-electron processes, which is the
subject of our work, there is another specific
enhancement at electron energies below 1 meV. This enhancement increases with
the charge of the ion, and is observed for all ions including fully stripped
ones, see Gao {\em et al.} 1997 and Uwira {\em et al.} 1997$b$, and we do not
consider it here.}. Its
magnitude ranges from a factor of about ten for Ar$^{13+}$ (Gao {\em et al.}
1995), Au$^{50+}$ and Pb$^{53+}$ (Uwira {\em et al.} 1997$a$), and U$^{28+}$
(M\"uller and Wolf 1997), to over a hundred for Au$^{25+}$ (Hoffknecht
{\em et al.} 1998). This enhancement is sensitive to
the electronic structure of the target, e.g., the recombination rates of
Au$^{49+}$ and Au$^{51+}$ are much smaller than that of Au$^{50+}$
(Uwira {\em et al.} 1997$a$). For few-electron ions, e.g., C$^{4+}$,
Ne$^{7+}$ and
Ar$^{15+}$ the observed rates are described well by the sum of the direct and
dielectronic recombination rates (Schennach {\em et al.} 1994,
Zong {\em et al.} 1997, Schuch {\em et al.} 1997). In more complicated cases,
like U$^{28+}$ or Au$^{25+}$, the questions of what are the particular
resonances just above the threshold and how they contribute
to the recombination ``remain a mystery'' (Mitnik {\em et al.} 1998).

\section{Spectrum and eigenstates of Au$^{24+}$}

\begin{figure}[t]
\vspace{8pt}
\hspace{-35pt}
\epsfxsize=11cm
\centering\leavevmode\epsfbox{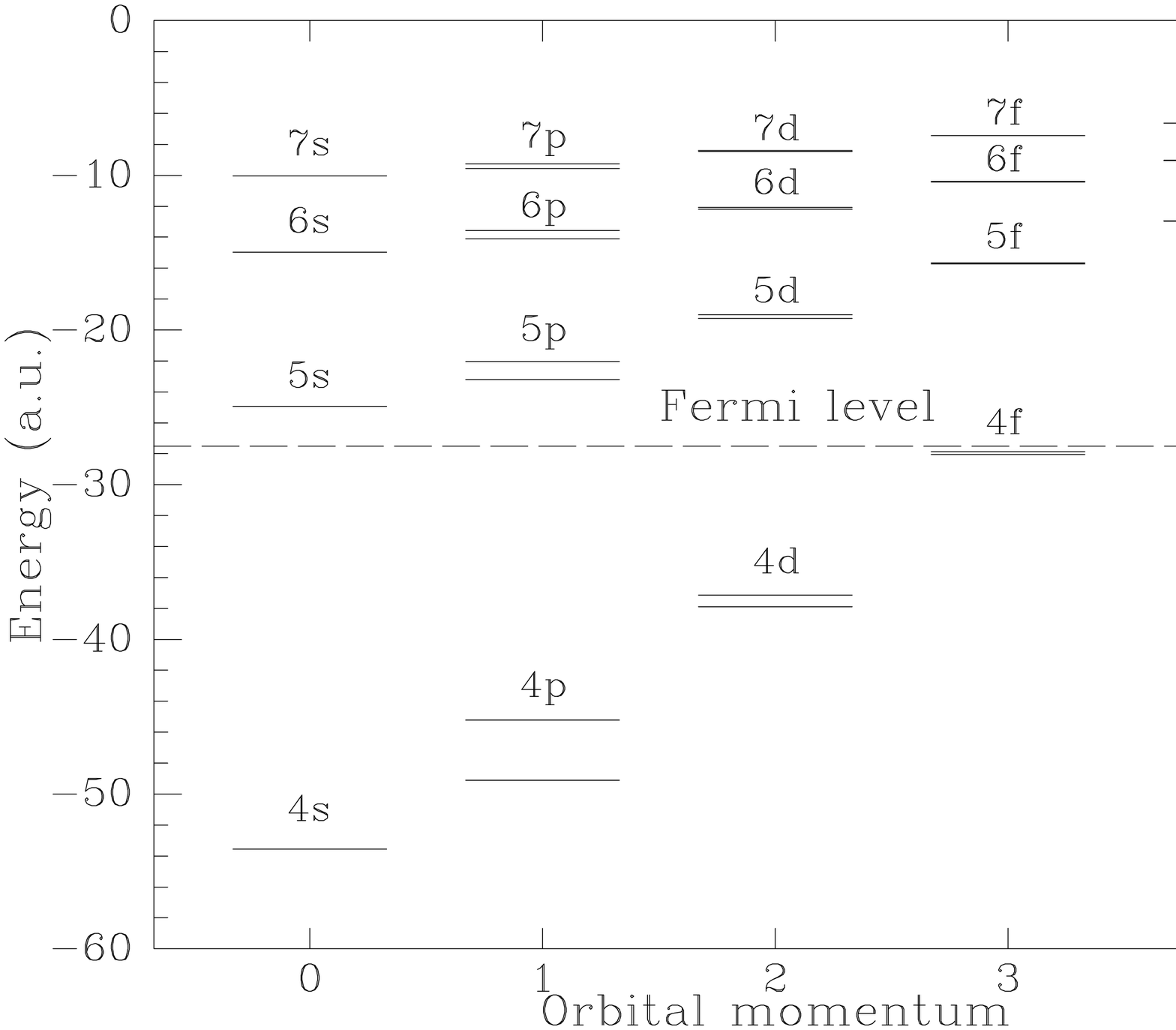}
\vspace{8pt}
\caption{Electron orbitals of Au$^{24+}$ from the relativistic Hartree-Fock
calculation.}
\label{fig:orb}
\end{figure}

Let us consider the problem of electron recombination on Au$^{25+}$. Due to
electron correlations the low-energy electron can be captured into an excited
state of the compound Au$^{24+}$ ion. This system is the main object of our
analysis. Au$^{24+}$ has 55 electrons. Its ground state belongs to the
$1s^2\dots 4f^9$ configuration. Figure \ref{fig:orb} shows the energies of
its relativistic orbitals $nlj$ obtained in the relativistic Hartree-Fock
calculation. All orbitals below the Fermi level, $1s$ to $4f$, were
obtained in the self-consistent calculation of the Au$^{24+}$ ground state.
Each of the excited orbitals above the Fermi level -- $5s$, $5p$, etc., was
calculated by placing one electron into it, in the field of the frozen
$1s^2\dots 4f^8$ core. The energy of the highest orbital
occupied in the ground state is $\varepsilon _{4f_{7/2}}=-27.9$ a.u. This
value gives an estimate of the ionization potential of Au$^{24+}$:
$I\approx |\varepsilon _{4f_{7/2}}|$. Our relativistic
configuration-interaction (CI) calculation of the ground states of
Au$^{24+}4f^9$ and Au$^{24+}4f^8$ shows that they are characterized by
$J=\frac{15}{2}$ and 6, and their total energies are $-18792.36$ and
$-18764.80$ a.u., respectively. Thus, the ionization threshold of Au$^{24+}$
is $I=27.56$ a.u.$=750$ eV, in agreement with Hoffknecht {\em et al.} 1998.

The excited states of the ion are generated by transferring one, two,
three, etc. electrons from the ground state into empty orbitals
above the Fermi level (Fig. \ref{fig:orb}), or into the partially occupied
$4f$ orbitals. We are  interested in the excitation
spectrum of Au$^{24+}$ near its ionization threshold. This energy (27.5 a.u.)
is sufficient to push up a few of the nine $4f$ electrons, and even excite
one or two electrons from the $4d$ orbital. However, the
preceding $4p$ orbital is already deep enough to be considered inactive.
Thus, we treat Au$^{24+}$ as a system of $n=19$ electrons above the frozen
Kr-like $1s^2\dots 4p^6$ core. Note also that infinite Rydberg series
corresponding to the excitations of one electron in the field of Au$^{25+}$
belong to the single-particle aspect of the Au$^{25+}$ $+$ $e^-$ problem, and
we do not consider them here.

The number of multielectron states obtained by distributing 19 electrons over
31 relativistic orbitals, $4d_{3/2}$ through to $7g_{9/2}$, is enormous, even
if we are only interested in the excitation energies below 27.5 a.u. It is
impossible to perform any CI calculation for them. However, there is another
simpler way to analyse the spectrum. The scale of the configuration interaction
strength is determined by the two-body Coulomb matrix elements which transfer
electrons between different configurations. Their typical
size in neutral atoms is $\sim 1 $ eV, and in Au$^{24+}$ it is about 1 a.u.,
which is roughly 25 times greater, due to the smaller radius of the ion.
This scale is much smaller than $I$. Configuration mixing aside, the CI does
not shift the mean energies of the configurations.
Therefore, we can construct the excitation spectrum of Au$^{24+}$ by
calculating the mean energies $E_i$ of the configurations, and the numbers of
many-electron states $N_k$ within each of them:
\begin{eqnarray}\label{eq:Ei}
E_i&=&E_{\rm core}+\sum _a\epsilon _an_a+
\sum _{a\leq b}\frac{n_a(n_b-\delta _{ab})} {1+\delta _{ab}}U_{ab}~,\\
N_i&=&\prod _a\frac{g_a!}{n_a!(g_a-n_a)!}~,\label{eq:Ni}
\end{eqnarray}
where $n_a$ are the integer orbital occupation numbers of the relativistic
orbitals in a given configuration ($\sum _a n_a=n$), $\epsilon _a=
\langle a|H_{\rm core}|a\rangle $ is the single-particle energy of the orbital
$a$ in the field of the core, $g_a=2j_a+1$, and $U_{ab}$ are the average
Coulomb matrix elements for the electrons in orbitals $a$ and $b$ (direct
minus exchange):
\begin{equation}\label{eq:Uab}
U_{ab}=\frac{g_a}{g_a-\delta _{ab}}\left[ R_{abab}^{(0)}-
\sum _\lambda \delta _p R_{abba}^{(\lambda )}
\left( {j_a\atop \frac{1}{2}} {j_b\atop -\frac{1}{2}} {\lambda \atop 0}
\right) ^2\right] .
\end{equation}
Here $R_{abba}^{(\lambda )}$ is the two-body radial Coulomb integral
of $\lambda $ multipole, and $\delta _p=1$ when $l_a+l_b+\lambda $ is even,
and 0 otherwise. The mean energy of the lowest configuration
$4d^{10}4f_{5/2}^64f_{7/2}^3$ obtained from Eq. (\ref{eq:Ei}) is just
0.28 a.u. above the CI ground state.

\begin{figure}[t]
\vspace{8pt}
\hspace{-35pt}
\epsfxsize=12.0cm
\centering\leavevmode\epsfbox{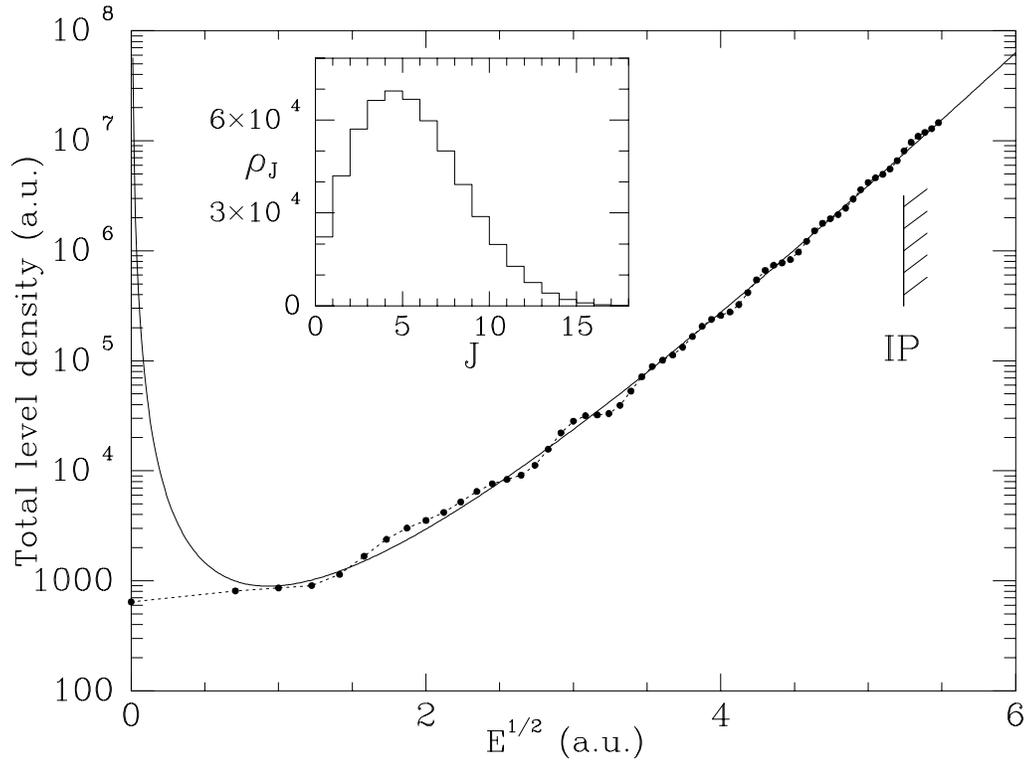}
\vspace{8pt}
\caption{Level density in Au$^{24+}$. Full circles connected by dotted line
to guide the eye is the result of our numerical calculation. Solid line is the
analytical fit, Eq. (\ref{eq:fit}). The inset shows the densities of states
with different $J$ near the ionization threshold $E= I$.}
\label{fig:dens}
\end{figure}

Using Eqs. (\ref{eq:Ei})--(\ref{eq:Uab}) we find that there are 9000
configurations within 35 a.u. of the Au$^{24+}$ ground state. They comprise a
total of $2.1\times 10^8$ many-electron states. If we allow for about 10
different values of $J$, $2J+1$ values of $J_z$ and the two parities,
there would still be about $5\times 10^5$ states in each $J^\pi $ manifold.
In Fig. \ref{fig:dens} we show the total density of states for Au$^{24+}$
as a function of $\sqrt{E}$, where $E$ is the excitation energy of the system
above the ground state. It is obtained by smoothing out the small-scale
fluctuations of the level density
\begin{equation}\label{eq:rho}
\rho (E)=\sum _iN_i\delta (E-E_i)
\end{equation}
by folding it with a Gaussian with $1$ a.u. variance. In reality this
averaging is done by the interaction and mixing of the configurations
(Flambaum {\em et al.} 1994), but the result is expected to be the same.
The inset on Fig. \ref{fig:dens} presents a break-up of the total density near
the ionization threshold into the densities of states with given $J$:
$\rho (E)=\sum _J(2J+1)\rho _J(E)$. The most abundant values are
$J=\frac{5}{2}$ to $\frac{15}{2}$. For a given parity the density of such
states at $E\approx I$ is $\rho _{J^\pi }\approx 3.5\times 10^4$ a.u.,
which corresponds to the mean level spacing $D=1/\rho _{J^\pi }\sim 1$ meV.
Figure \ref{fig:dens} demonstrates the
characteristic $\rho \propto \exp (a\sqrt{E})$ behaviour of the level density
predicted by the Fermi-gas model (Bohr and Mottelson 1969), where $a$ is
related to the single-particle level density at the Fermi level
$g(\varepsilon _F)$ as $a=[2\pi ^2 g(\varepsilon _F)/3] ^{1/2}$.
$g(\varepsilon _F)=3a^2/2\pi^2$. We obtain an accurate fit
of the level density at $E>1$ a.u. by using a Fermi-gas model anzats
\begin{equation}\label{eq:fit}
\rho (E)=AE^{-\nu }\exp (a\sqrt{E})~,
\end{equation}
with  $A=31.6$, $\nu =1.56$, and $a=3.35$. The corresponding value of
$g(\varepsilon _F)=3a^2/2\pi^2=1.7$ a.u. is close to what one obtains from
the Hartree-Fock orbital spectrum in Fig. \ref{fig:orb}. The other two
parameters are different from the non-interacting Fermi-gas model values
$A=1/\sqrt{48}$ and $\nu =1$. The latter values in fact lead to strong
underestimation of the level density. For most abundant $J ^\pi$ states the
density $\rho _{J^\pi }(E)$ is given by Eq. (\ref{eq:fit}) with
$A_{J^\pi }\approx 0.15$.

At first sight the huge level density makes the spectrum of Au$^{24+}$
enormously complicated. On the other hand, this complexity enables one to
analyse the system using statistical methods. The interaction between
multiply-excited configuration states mixes them completely, and they
loose their individual features. In this regime the spectral statistics become
close to those of a random matrix ensemble, the eigenstates cannot be
characterized by any quantum numbers except the exact ones (energy and
$J^\pi $), and the orbital occupation numbers deviate prominently from
integers. This regime can be described as {\em many-body quantum chaos}. We
have extensively studied it in direct numerical calculations for the rare-earth
atom of Ce -- a system with four valence electrons (Flambaum {\em et al.} 1994,
1996, 1998$a$, 1998$b$, Gribakina {\em et al.} 1995).

The strength of the configuration mixing is characterized by the spreading
width $\Gamma _{\rm spr}$. For a configuration basis state $\Phi _k$ with
energy $E_k$ it defines the energy range $|E-E_k|\lesssim \Gamma _{\rm spr}$
of eigenstates in which this basis state noticeably participates. By the same
token it shows that a particular eigenstate $\Psi =\sum _kC_k\Phi _k$ contains
a large number $N\sim \Gamma _{\rm spr}/D$ of {\em principal components} --
basis states characterized by $C_k\sim 1/\sqrt{N}$. Outside the spreading width
$C_k$ decrease. This effect is usually referred to as {\em localization}.
Apart from this, $C_k$ behave closely to Gaussian random variables
(Flambaum {\em et al.} 1994). The effect of spreading is approximated well by
the Breit-Wigner shape of the mean-squared components (Bohr and Mottelson 1969)
\begin{equation}\label{eq:BW}
\overline {C_k^2}(E)=\frac{1}{N}\,\frac{\Gamma _{\rm spr}^2/4}
{(E_k-E)^2+\Gamma _{\rm spr}^2/4}~.
\end{equation}
The normalization $\sum _k\overline {C_k^2}=1$ yields $N=\pi\Gamma /2D$.
In systems with small level spacings $D$ the number of principal components
$N$ can be very large. It reaches several hundreds in Ce, and can be as large
as $10^6$ in complex nuclei. At $|E_k-E|>\Gamma $ Eq. (\ref{eq:BW})
gives $\overline {C_k^2}(E)\propto 1/(E_k-E)^2$, which corresponds to the
simple first-order perturbation theory dependence with constant mean-squared
mixing matrix elements. In real systems the mixing between distant (in the
sense of their unperturbed energies) basis states is usually suppressed.
Accordingly, the Hamiltonian matrix is characterized by certain bandedness,
i.e., the off-diagonal matrix elements $H_{ij}$ decrease as one moves away
from the main diagonal $i=j$ (Gribakina {\em et al.} 1995). This causes a
faster, close to exponential, decrease of the mean-squared components at
large $|E_k-E|$ (Flambaum {\em et al.} 1994).

\begin{figure}[t]
\vspace{8pt}
\hspace{-35pt}
\epsfxsize=15.0cm
\epsfysize=12cm
%\centering\leavevmode\epsfbox{Aufig3.eps}
\epsfbox[-150 50 624 646]{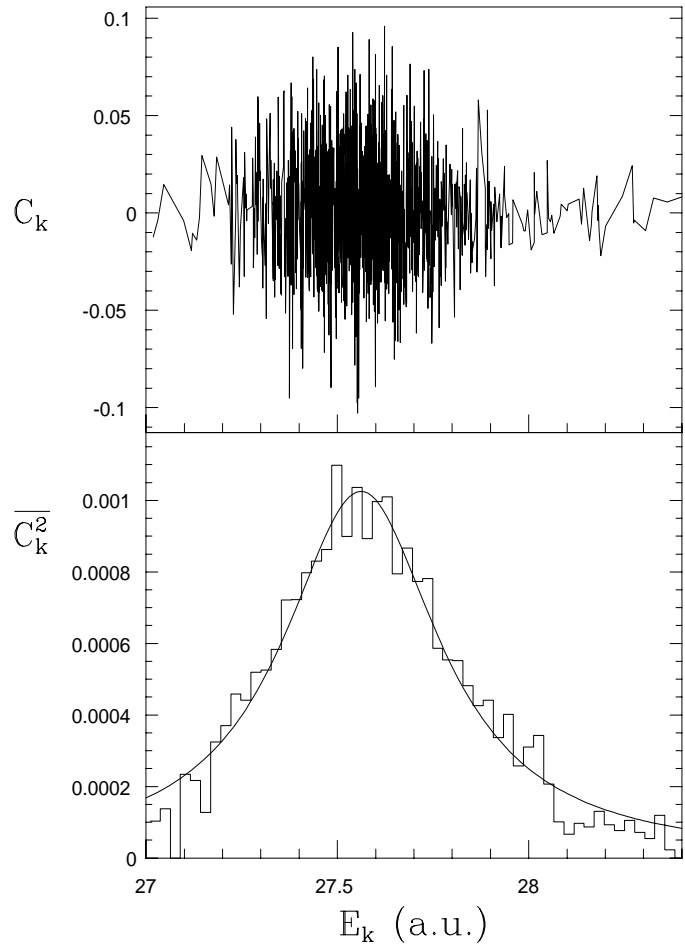}
\vspace{8pt}
\caption{Components of the 590th $J^\pi =\frac{13}{2}^-$ eigenstate from a
two-configuration calculation (top), and a fit of
$\protect \overline{C_k^2}(E)$ by the Breit-Wigner formula (\ref{eq:BW})
(bottom).}
\label{fig:comp}
\end{figure}

\begin{figure}[t]
\hspace{-35pt}
\epsfxsize=15.0cm
\centering\leavevmode\epsfbox{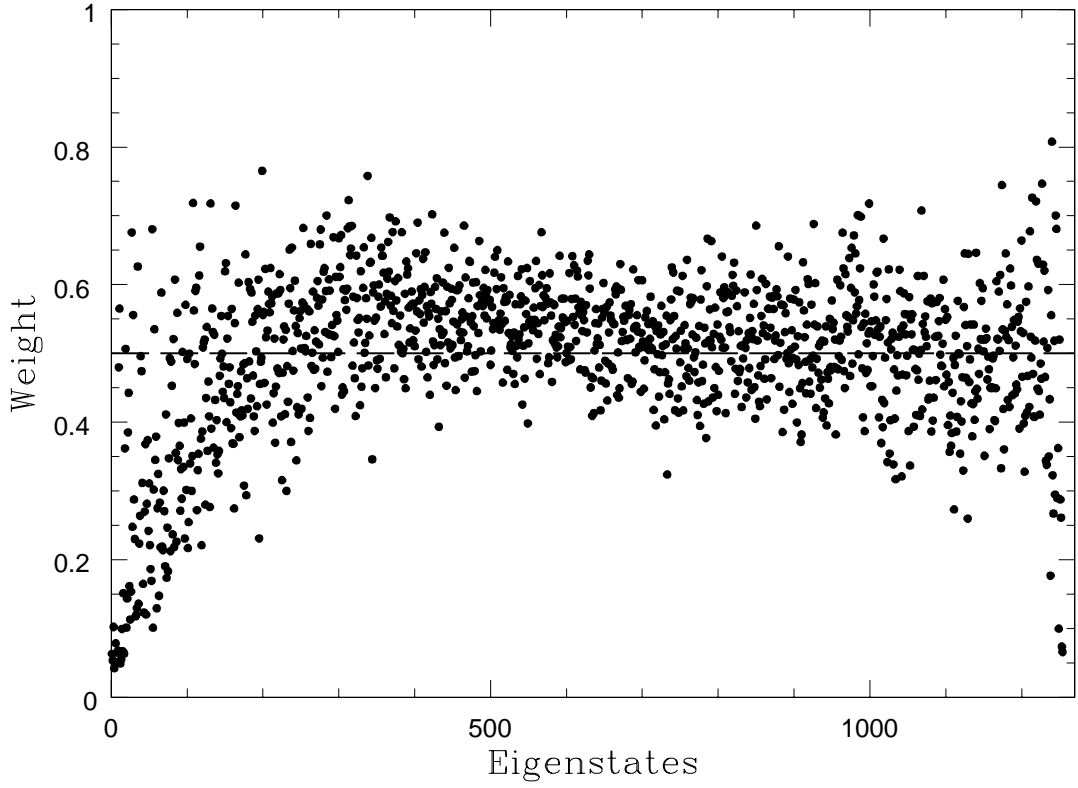}
\vspace{-50pt}
\caption{Weights of the $4f_{5/2}^34f_{7/2}^35p_{1/2}5p_{3/2}5f_{7/2}$
configuration in the $J^\pi =\frac{13}{2}^-$ eigenstates obtained in the
two-configuration calculation. Note that for a few lower eigenstates the
weights of this configuration are small, because its mean energy is about
0.03 a.u. higher than that of the other configuration.}
\label{fig:weight}
\end{figure}

In Fig. \ref{fig:comp} we illustrate the behaviour of the eigenstate components
by the results of a CI calculation which includes just two odd configurations
of Au$^{24+}$ with energies close to the ionization threshold:
$4f_{5/2}^34f_{7/2}^35p_{1/2}5p_{3/2}5f_{7/2}$ and
$4f_{5/2}^34f_{7/2}^35p_{1/2}5d_{3/2}5g_{7/2}$. These two configurations
produce a total of 143360 many-electron states with $J$ ranging from
$\frac{1}{2}$ to 17.5. As an example we present the results obtained
by diagonalization of the Hamiltonian matrix for $J^\pi =\frac{13}{2}^-$.
This total angular momentum value is among the most abundant in the spectrum,
as there are 1254 $J^\pi =\frac{13}{2}^-$ states. The mixing of the two
configurations included is practically complete, since the weight of each
configuration in every eigenstate is close to 50\%, Fig. \ref{fig:weight}.
Shown in the upper part
of Fig. \ref{fig:comp} are the components of the 590th eigenstate from the
middle of the two-configuration spectrum. Both the fluctuations of
$C_k$ as function of basis state $k$, and the localization of the eigenstate
components in the vicinity of the corresponding eigenvalue ($E=27.51$ a.u.
above the Au$^{24+}$ ground state) are evident.

A Breit-Wigner fit of the mean-squared components yields $N=975$ and
$\Gamma _{\rm spr}=0.50$ a.u., see lower part of Fig. \ref{fig:comp}. When the
calculations are performed for one of the above configurations, $N$ is about
two times smaller, but $\Gamma _{\rm spr}$ is practically the same. The
spreading width is related to
the mean-squared off-diagonal Hamiltonian matrix element and the mean level
spacing as $\Gamma _{\rm spr}\simeq 2\pi \overline{H_{ij}^2}/D$ (Bohr and
Mottelson 1969).
It is known to be a robust characteristic of the system.
When more configurations are included, both $D$ and $\overline{H_{ij}^2}$
decrease, whereas $\Gamma _{\rm spr}$ does not change much. If one could do a
full-scale CI calculation near the ionization threshold of Au$^{24+}$ one
would obtain eigenstates with
$N=(\pi /2)\Gamma _{\rm spr} \rho _{J^\pi}\sim 3\times 10^4$ principal
components.

\section{Statistical approach}

\begin{figure}[t]
\hspace{-100pt}
\epsfysize=17cm
\epsfbox[-150 0 624 736]{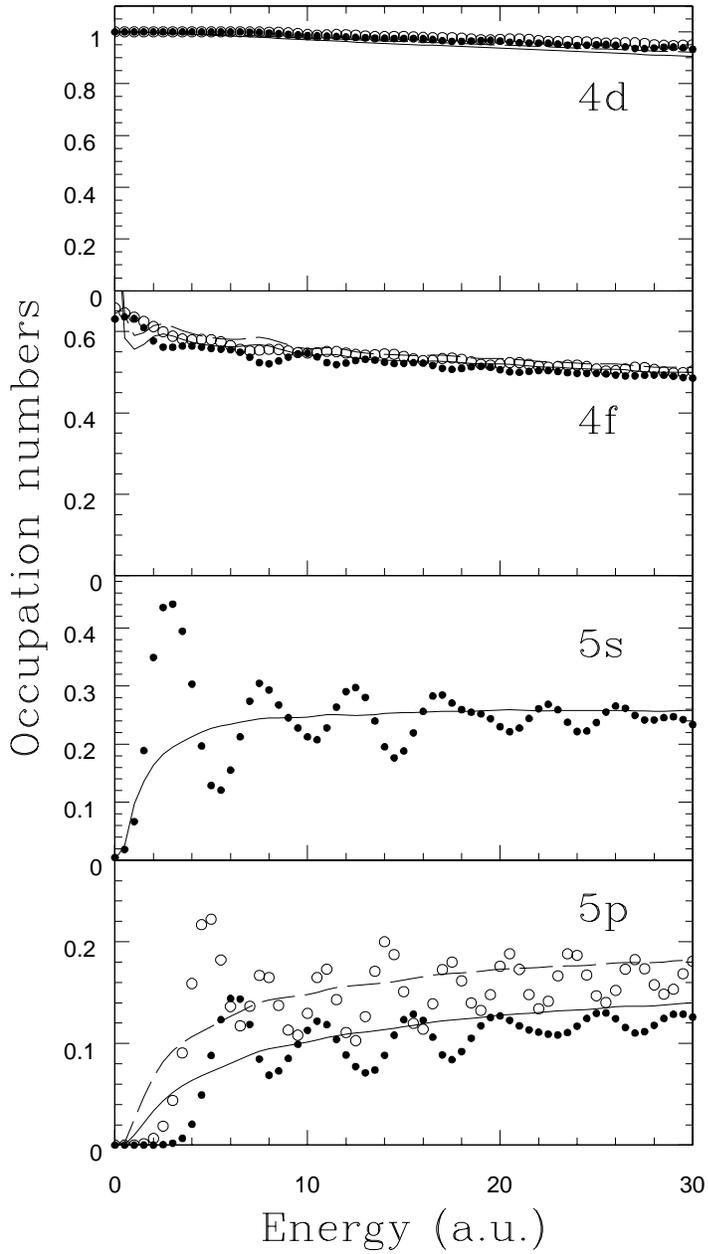}
%\centering\leavevmode\epsfbox{AuAJPf4.eps}
\vspace{0pt}
\caption{Energy dependence of the single-particle occupation numbers
$n_a/g_a$ calculated numerically from Eq. (\ref{eq:na}) for $4d$, $4f$, $5s$
and $5p$ orbitals: solid and open circles correspond to $j=l\pm \frac{1}{2}$
subshells,
respectively. Solid and dashed lines ($j=l\pm \frac{1}{2}$, respectively)
show the results obtained from the FD formula using the energy-dependent
orbital energies and chemical potential, and the canonical relation between
the temperature and the excitation energy (solid line in Fig. \ref{fig:temp}).}
\label{fig:occup}
\end{figure}

The spreading of the basis states due to configuration interaction introduces
natural statistical averaging in the system. Based on this averaging, a
statistical theory of finite Fermi systems of interacting particles can be
developed (Flambaum and Izrailev 1997$a$, 1997$b$). It enables one to calculate
various properties
of the system  as sums over the basis states, without actually diagonalizing
the Hamiltonian matrix. For example, the mean orbital occupations numbers can
be obtained as
\begin{equation}\label{eq:na}
n_a(E)=\sum _k\overline{C_k^2}(E)n_a^{(k)}
\end{equation}
where $n_a^{(k)}$ is the occupation number of the orbital $a$ in the basis
state $k$. To demonstrate how it works we have used a simple Gaussian model
spreading
\begin{equation}\label{eq:Gauss}
\overline{C_k^2}(E)\propto \exp \left[ -\frac{(E_k-E)^2}{2\Delta _E^2}\right]
\end{equation}
and calculated the mean orbital occupation numbers as functions of the
excitation energy $E$ using $\Delta _E=1$ a.u., Fig. \ref{fig:occup}. Of
course, in our calculation we sum over the configurations, rather than the
actual many-electron basis states, and use their mean energies and weights
given by Eqs. (\ref{eq:Ei}) and (\ref{eq:Ni}), cf. Eq. (\ref{eq:rho}).

The oscillatory dependence with the period of about 3--4 a.u. is due to the
shell structure of the Au$^{24+}$ ion, Fig. \ref{fig:orb}. As the excitation
energy increases the oscillations die out. Apart from this the occupation
numbers of the orbitals below the Fermi level ($4d$ and $4f$) decrease, and
those above it ($5s$, $5p$, etc.) increase, as one would expect in a Fermi
system. It seems very natural to try to describe this behaviour in the spirit
of statistical
mechanics, by introducing temperature and applying the standard Fermi-Dirac
(FD) distribution (Flambaum {\em et al.} 1998$b$). Temperature has long
been used to describe highly excited nuclei, and the question of
thermalization was investigated recently in numerical calculations for the
$s-d$ shell nuclear model (Horoi {\em et al.} 1995, Zelevinsky {\em et al.}
1996). Of course, temperature can
always be used to describe the equilibrium of a macroscopic system that
contains a large number of particles, or to describe a small system interacting
with a heat bath. In what follows we are going to see if the notion of
temperature can be applied to our isolated system with a small number
of active particles. The total number of electrons in Au$^{24+}$ is quite
large, however most of them are {\em inactive} at the excitation
energies at or below the ionization threshold.

The formula for the single-particle occupation numbers $\nu _a=n_a/g_a$
($0\leq \nu _a\leq 1$)
\begin{equation}\label{eq:FD}
\nu _a=\frac{1}{1+\exp [(\varepsilon _a-\mu )/T] }~,
\end{equation}
at a given temperature $T$ and chemical potential $\mu $ depends on the
single-particle orbital energies $\varepsilon _a$. These energies are well
defined for non-interacting particles in a given potential. For interacting
particles (electrons in an atom or ion) one can introduce single-particle
orbitals and energies using a {\em mean field} approximation, e.g. the
Hartree-Fock method. From this points of view we could use the orbital
energies $\varepsilon _a^{\rm HF}$ found in the mean field of the Au$^{24+}$
ground state $1s^2\dots 4d^{10}4f^9$, Fig. \ref{fig:orb}. However, they may
only be suitable at low excitation energies, when the mean field is close to
that of the ground-state Au$^{24+}$.

\begin{figure}[t]
\vspace{8pt}
\hspace{-55pt}
\epsfxsize=12cm
\centering\leavevmode\epsfbox{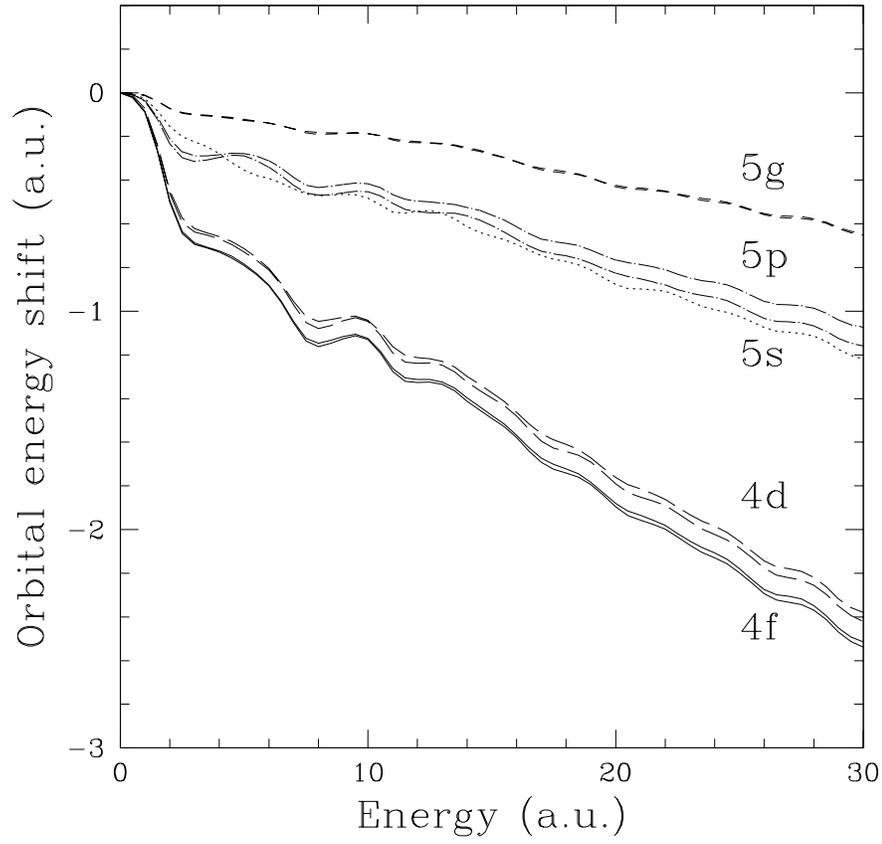}
\vspace{8pt}
\caption{Shifts of the single-particle orbital energies, Eq. (\ref{eq:deleps}),
as functions of the excitation energy for the $4d_{3/2,5/2}$ (long dash),
$4f_{5/2,7/2}$ (solid), $5s$ (dotted), $5p_{1/2,3/2}$ (chain), and
$5g_{7/2,9/2}$ (short dash) orbitals.}
\label{fig:orbiten}
\end{figure}

As the excitation energy increases the orbital occupation numbers change
noticeably, as shown by Fig. \ref{fig:occup}. This gives rise to a change of
the mean field, and as a result, the orbital energies are shifted by
\begin{equation}\label{eq:deleps}
\delta \varepsilon _a(E)=\sum _bU_{ab}\delta n_b(E)~,
\end{equation}
where $\delta n_b=n_b(E)-n_b(0)$ is the difference between the occupation
numbers at energy $E$ and in the ground state at $E=0$. Using our numerical
energy-dependent occupation numbers we find the energy dependence of the
orbital energies, shown for a few low-lying orbitals in Fig.
\ref{fig:orbiten}. With the increase of the excitation energy the electrons are
transferred into higher orbitals which have larger radii. Accordingly,
the electron cloud becomes more diffuse, the screening of the nuclear potential
is reduced, and the orbital energies go down. This effect is especially
strong for the inner $4d$ and $4f$ orbitals. As we will see below the
shift of the lower orbital energies is comparable with the temperature of the
system, and it has to be taken into account when applying the FD formula.

\begin{figure}[t]
\hspace{-100pt}
\epsfysize=17cm
\epsfbox[-150 0 624 696]{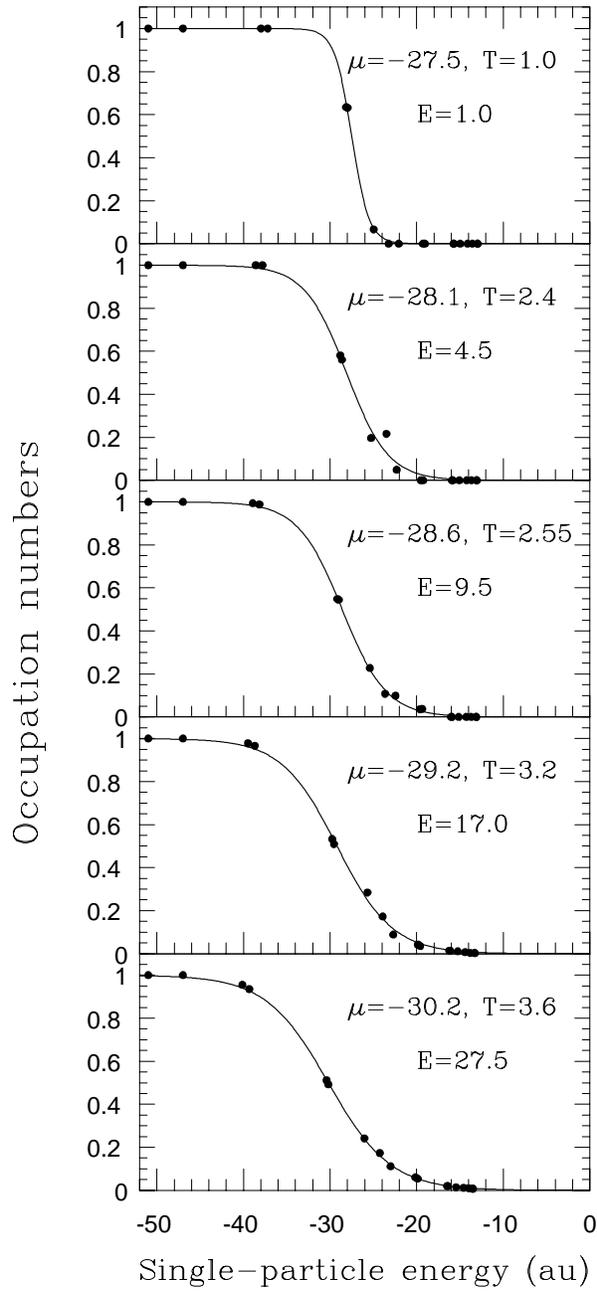}
%\centering\leavevmode\epsfbox{AuAJPf6.eps}
\vspace{0pt}
\caption{Orbital occupation numbers in Au$^{24+}$ calculated numerically
from Eq. (\ref{eq:na}) at excitation energies $E=1$, $4.5$, $9.5$, 17 and
$27.5$ a.u. (solid circles), and the Fermi-Dirac distributions (solid line)
with temperature $T$ and chemical potential $\mu $ chosen to give best
fits of the numerical data.}
\label{fig:FD}
\end{figure}

\begin{figure}[t]
\vspace{8pt}
\hspace{-55pt}
\epsfxsize=8cm
\centering\leavevmode\epsfbox{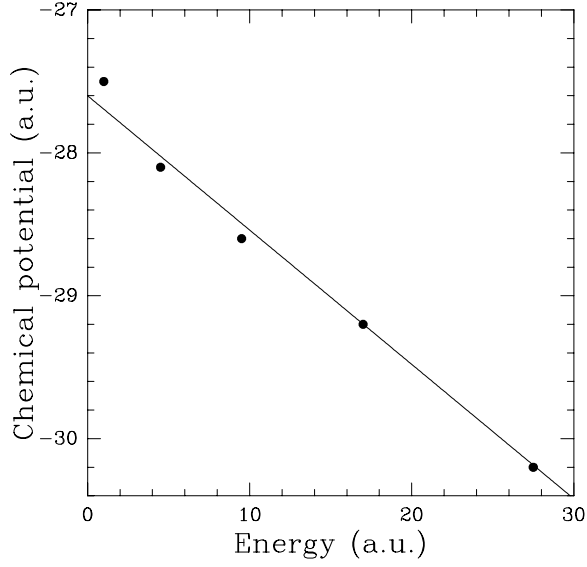}
\vspace{8pt}
\caption{Chemical potential obtained from the FD distribution fits of the
occupation numbers, Fig.~\ref{fig:FD}, as a function of the excitation energy
of Au$^{24+}$ (solid circles). Solid line is a simple linear fit
$\mu =-27.6-0.094E$ a.u.}
\label{fig:mu}
\end{figure}

\begin{figure}[t]
\vspace{8pt}
\hspace{-55pt}
\epsfxsize=7cm
\centering\leavevmode\epsfbox{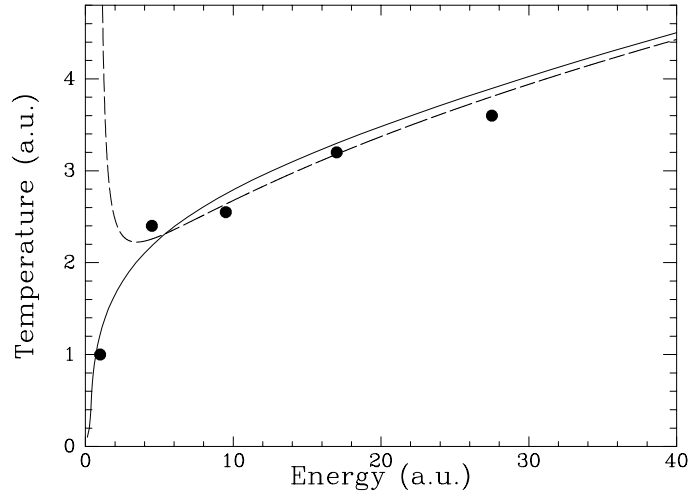}
\vspace{8pt}
\caption{Temperature vs energy for Au$^{24+}$. Solid line - canonical
definition, Eq. (\ref{eq:ET}); dashed line - statistical physics definition,
Eq. (\ref{eq:T}), which uses the density fit (\ref{eq:fit}); solid circles --
Fermi-Dirac fits of the occupation numbers.}
\label{fig:temp}
\end{figure}

In Fig. \ref{fig:FD} we present the single-particle occupation numbers
at five different excitation energies, as functions of the
shifted orbital energies
\begin{equation}\label{eq:epsE}
\varepsilon _a(E)=\varepsilon _a(0)+ \delta \varepsilon _a(E)~,
\end{equation}
where we take $\varepsilon _a(0)=\varepsilon _a^{\rm HF}$. The numerical values
agree well with the FD distribution Eq. (\ref{eq:FD}), obtained
by the least-square fits of the temperature $T$ and chemical potential $\mu $
(solid lines). Figures \ref{fig:mu} and \ref{fig:temp} present the dependence
of $\mu $ and the ``Fermi-Dirac temperature'' $T$ on the energy of the system
(solid circles).

Occupation numbers aside, the relation between the temperature and energy can
be defined by the level density $\rho (E)$, Eq. (\ref{eq:rho}), through the
canonical average
\begin{equation}\label{eq:ET}
E(T)=\frac{\int e^{-E/T}E\rho (E)dE}{\int e^{-E/T}\rho (E)dE}=
\frac{\sum _i E_i N_i e^{-E_i/T}}{\sum _i N_i e^{-E_i/T}}~,
\end{equation}
or from the statistical physics formula
\begin{equation}\label{eq:T}
T^{-1}=\frac{d\ln [\rho (E)]}{dE}~,
\end{equation}
using the smooth fit (\ref{eq:fit}). The latter yields $T\simeq 2\sqrt{E}/a$,
or $E\propto T^2$, characteristic of the Fermi systems. Figure \ref{fig:temp}
shows that for the energies above 3 a.u. all three definitions of temperature
give close values. As is known, the expansion of the chemical potential in a
Fermi system at small temperatures starts with a $T^2$ term (Landau and
Lifshitz 1969).
Accordingly, its shift from the ground-state value is proportional to the
energy. Indeed, a simple linear fit $\mu =-27.6-0.094E$ closely follows the
numerical values in Fig. \ref{fig:mu}.

If we use $T(E)$ given by the canonical definition and the linear fit of
$\mu $, together with the orbital energies [Eq. (\ref{eq:epsE})], the FD
formula gives smooth energy dependencies of the occupation numbers,
see Fig. \ref{fig:occup}. They reproduce the behaviour of the numerical
occupation numbers averaged over the shell-structure fluctuations.

\section{Direct and resonant recombination}

Let us now estimate the direct and resonant contributions to the recombination
rate of Au$^{25+}$. The direct radiative recombination cross section is
estimated by introducing an effective ionic charge $Z_i$ into the Kramers
formula, which describes radiative electron capture in the Coulomb potential,
see e.g. Sobelman 1992,
\begin{equation}\label{eq:Kramers}
\sigma _n ^{\rm (d)}=\frac{32\pi }{3\sqrt{3}c^3}\,\frac{(Z_i^2{\rm Ryd})^2}
{n \varepsilon (Z_i^2{\rm Ryd} +n^2\varepsilon )}~,
\end{equation}
where $\varepsilon $ is the initial electron energy, $n$ is the principal
quantum number of the final electron state, and atomic units are used
(${\rm Ryd}=\frac{1}{2}$ a.u.). If we are interested in the total
recombination cross section the sum over $n$ must be calculated,
\begin{equation}\label{eq:sumsig}
\sigma ^{\rm (d)}=\sum _n \sigma _n ^{\rm (d)}.
\end{equation}
Due to the $n^{-1}$ factor in Eq. (\ref{eq:Kramers}) this sum diverges
logarithmically, until values of $n\sim n_{\rm max}$ are reached,
where $n^2_{\rm max}\varepsilon =Z_i^2{\rm Ryd}$, after which it converges
rapidly. With the logarithmic accuracy the result is given by
\begin{equation}\label{eq:sigmad}
\sigma ^{\rm (d)}\simeq \frac{32\pi }{3\sqrt{3}c^3}\,\frac{\rm Ryd}
{\varepsilon }
Z_i^2\ln \left( \frac{Z_i}{n_0}\sqrt{\frac{\rm Ryd}{\varepsilon }}\right)~,
\end{equation}
where $n_0$ is the principal quantum number of the lowest unoccupied ionic
orbital, which determines the lower limit in the summation over $n$. Using
$Z_i=25$, $n_0=5$ for electron recombination with Au$^{25+}$, and choosing a
small electron energy of $\varepsilon =0.1$ eV we obtain
$\sigma ^{\rm (d)}\approx 7\times 10^{-17}$ cm$^{2}$. This  corresponds to
the recombination rate of $\lambda =\sigma v= 1.3\times 10^{-9}$
cm$^3$s$^{-1}$, which is two orders of magnitude smaller than the experimental
$\lambda =1.8\times 10^{-7}$ cm$^3$s$^{-1}$ at this energy
(Hoffknecht {\em et al.} 1998).

The electron energy of 0.1 eV is equal to the transversal temperature of the
electron beam in the experiment, whereas the longitudinal temperature is
much smaller, 1 meV. Therefore, to make estimates of the recombination rates
at this and higher energies one can use the the cross sections without
averaging over the Maxwellian velocity distribution.
It is also important that the energy dependence of the experimental
recombination rate is in agreement with that of the direct radiative capture
for electron energies 1 meV$<\varepsilon \lesssim 1$ eV. The latter is
basically
given by the $1/\varepsilon $ factor in Eq. (\ref{eq:sigmad}). The experimental
data of Hoffknecht {\em et al.} (1998) is reproduced well by the direct
radiative rate multiplied by a factor of 150.

The cross section of resonant radiative capture averaged over the
resonances is (Landau and Lifshitz 1977)
\begin{equation}\label{eq:sigmar}
\sigma ^{\rm (r)}=\frac{\pi ^2}{\varepsilon }\,\frac{\Gamma _\gamma \Gamma _e}
{D(\Gamma _\gamma +\Gamma _e)}\approx \frac{\pi ^2}{\varepsilon }\,
\frac{\Gamma _\gamma}{D}\qquad (\Gamma _e\gg \Gamma _\gamma ),
\end{equation}
where $\Gamma _\gamma $ and $\Gamma _e$ are the mean radiative and
autoionization (or elastic) widths of the resonances, $D$ is the mean
resonance spacing, and we drop the statistical weights of the initial and
intermediate ionic states. The relation $\Gamma _e\gg \Gamma _\gamma $ is
usually valid for a few lower partial waves, where the electron interaction
is stronger than the electromagnetic one. Equation (\ref{eq:sigmar})
is written for the electron $s$-wave, and the contributions of higher
electron partial wave contain an extra factor $(2l+1)$.

The radiative width of the resonant state at energy $E\approx I$ is found by
summing the partial widths for all lower-lying states $E'=E-\omega $,
\begin{equation}\label{eq:width}
\Gamma _\gamma \approx \frac{3}{2J+1}\int _0^I\frac{4\omega ^3|d_\omega |^2}
{3c^3}\rho _{J^\pi }(I-\omega )d\omega ~,
\end{equation}
where the factor 3 accounts for $J'=J,~J\pm 1$, and $d_\omega $ is the reduced
dipole matrix element between the many-electron states. Because of the chaotic
structure of these states $d_\omega $ is suppressed compared to the typical
single-particle matrix element $d_0$: $d_\omega \sim d_0/\sqrt{N}$
(Flambaum and Sushkov 1984$a$, Flambaum {\em et al.} 1994, 1996). This estimate
for systems with
dense chaotic spectra in fact follows from the dipole sum rule: the number of
lines in the spectrum is large, $\propto D^{-1}\propto N$, consequently, the
line strengths are small, $|d_\omega |^2\sim |d_0|^2N^{-1}$.

The integrand in Eq. (\ref{eq:width}) peaks strongly because of the competition
between the $\omega ^3$ factor, and the level density
$\rho _{J^\pi }(I-\omega )$ that drops quickly as we go down from the
threshold, see Eq. (\ref{eq:fit}). As a result, the integral can be evaluated
by the saddle-point method. Using the statistical estimate of $d_\omega$ we
obtain
\begin{equation}\label{eq:sigres1}
\sigma ^{\rm (r)}=\frac{8\pi d_0^2}{(2J+1)c^3\varepsilon \Gamma _{\rm spr}}
\sqrt{\frac{2\pi}{3}}\rho _{J^\pi}(I-\omega _0)\omega _0^4~,
\end{equation}
where $\omega _0=6\sqrt{I}/a$ corresponds to the maximum of the decay
photon spectrum in Eq. (\ref{eq:width}). This cross section has the same energy
dependence as $\sigma ^{\rm (d)}$. Hence, it is also in agreement with the
energy dependence observed in the experiment, and we can estimate its
magnitude at one particular electron energy, e.g., 0.1 eV.
To do this we use a simple estimate of the single-particle dipole matrix
elements in the ion with charge $Z_i$: $d_0\sim Z_i^{-1}$, together with
$2J+1\approx 10$, and substitute $\Gamma _{\rm spr}=0.5$, 
$\omega _0=9.4$, and $\rho _{J^\pi}(I-\omega _0)=2.5\times 10^3$ a.u.
into Eq. (\ref{eq:sigres1}).
At $\varepsilon =0.1$ eV this gives $\sigma ^{\rm (r)}= 7\times 10^{-16}$
cm$^2$, therefore, $\sigma ^{\rm (r)}/\sigma ^{\rm (d)}= 10$, and we
obtain a factor of ten enhancement over the direct recombination due to
radiative capture into multiply excited resonant states (the corresponding
radiative width is $\Gamma _\gamma =2\times 10^{-7}$ a.u.). It comes from the
large effective number of final states in the radiative width in
Eq. (\ref{eq:width}) (numerically $\Gamma _\gamma \sim 2\times 10^{-7}$ a.u.).
This enhancement has been obtained for the electron $s$-wave. The
contributions of higher electron partial waves are similar to
Eq. (\ref{eq:sigmar}) times $(2l+1)$. Therefore, a few lower partial waves
($s$, $p$, $d$) produce resonant cross section values $10^2$ times greater
than $\sigma ^{\rm (d)}$, which matches the experimentally observed values.
With the increase of the orbital angular momentum $l$ of the electron the
capture width $\Gamma _e$ becomes smaller than the radiative width, and the
contribution of the higher partial waves to the resonant cross section is
suppressed.

\section{Conclusions}

In summary, the resonant radiative capture mechanism fully explains the
strongly enhanced recombination rates observed for eV electrons on
multicharged ions. Its origin is in the high level densities of chaotic
multiply-excited electron states in multicharged ions. The size of the
enhancement is sensitive to the electron structure of the ion, which
determines the level density. We have shown that a statistical approach can be
applied to the analysis of this complex system. One can also use a statistical
theory to calculate mean-squared matrix elements between multiply
excited chaotic states in terms of single-particle amplitudes, occupation
numbers, $\Gamma _{\rm spr}$ and $D$ (Flambaum and Vorov 1993,
Flambaum {\em et al.} 1994, 1996), and obtain accurate quantitative
information about the processes involving chaotic states and resonances.
At higher electron energies the resonant capture proceeds via so-called doorway
states (Bohr and Mottelson 1969) -- simple dielectronic autoionizing states,
which are then ``fragmented'' into the dense spectrum of multiply-excited
resonances (see Mitnik {\em et al.} 1998, Flambaum {\em et al.} 1996 and
Refs. therein).

\bibliographystyle{prsty}
% \bibliography{whole}

%*****************************************************************************
\figure

\end{document}